\documentclass[12pt]{iopart}
\usepackage{graphicx}
\usepackage{iopams}
\usepackage[caption=false]{subfig}
\usepackage{dcolumn}
\usepackage{bm}
\providecommand{\Rey}{\mbox{\ensuremath{\mathcal{R}}}}
\newcommand{\hidden}[1]{}

\begin{document}

\title[Quantum Analogs of Classical Wakes]{Quantum Analogs of Classical Wakes in Bose-Einstein Condensates}

\author{G. W. Stagg, N. G. Parker and C. F. Barenghi}

\address{Joint Quantum Centre (JQC) Durham-Newcastle, School of Mathematics and Statistics, Newcastle University, Newcastle upon Tyne, NE1 7RU, UK}
\ead{george.stagg@ncl.ac.uk}

\date{\today}

\begin{abstract}
We show that an elliptical obstacle moving through a Bose-Einstein condensate 
generates wakes of quantum vortices which resemble those of classical viscous flow past a cylinder or sphere.  The role of ellipticity is to facilitate the interaction of the vortices nucleated by the obstacle.  Initial steady symmetric wakes lose their symmetry and form clusters of like-signed vortices, in analogy to the classical B\'enard--von K\'arm\'an vortex street.  Our findings, demonstrated numerically in both two and three dimensions, confirm the intuition that a sufficiently large number of quanta of circulation reproduce classical
physics.
\end{abstract}

\maketitle

\section{Introduction}
\begin{figure}[b]
(a) \hspace{7cm}  (b) 
\\
\vspace{-0.4cm}

  \includegraphics[width=0.5\textwidth]{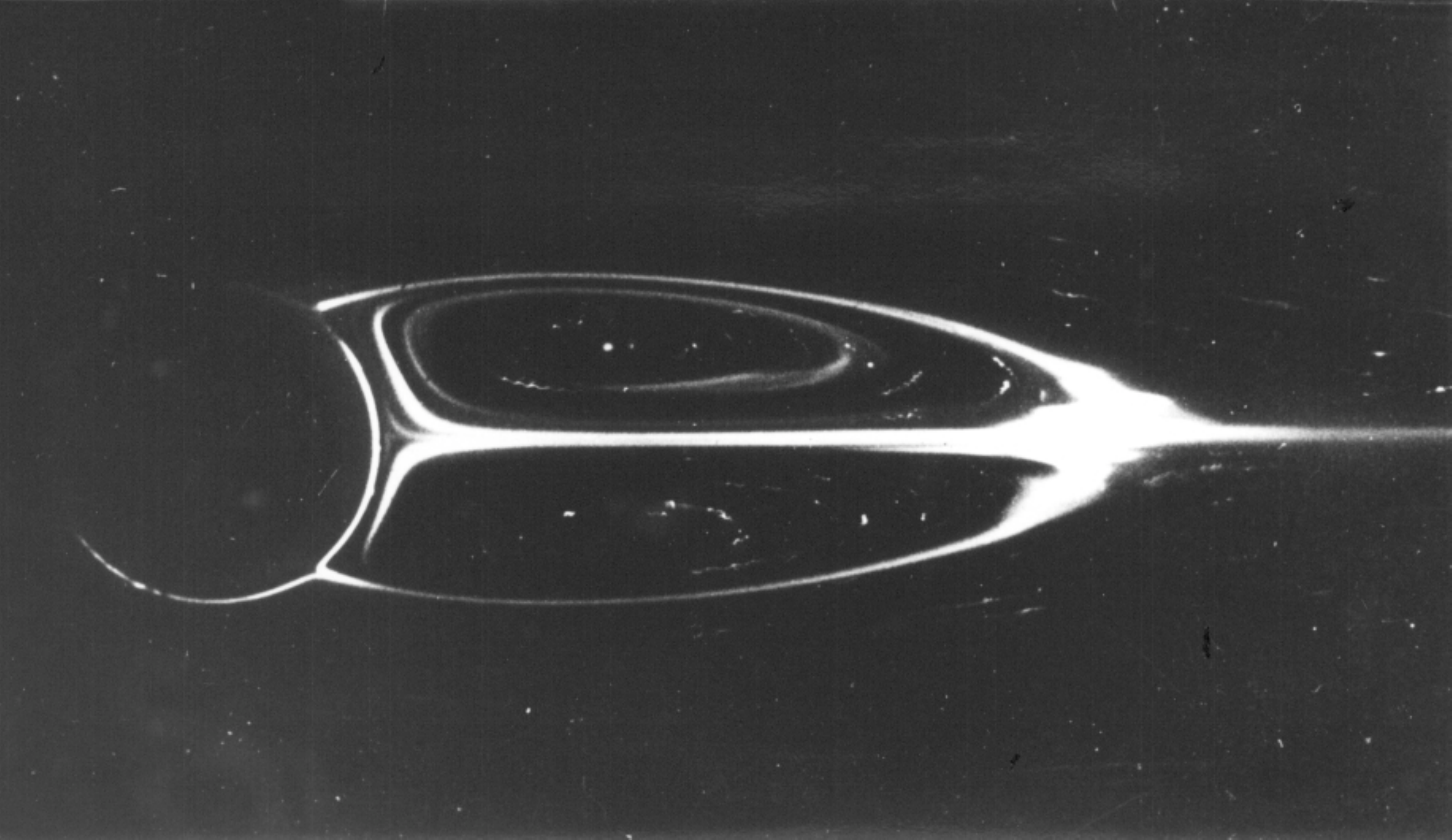}
  \includegraphics[width=0.5\textwidth]{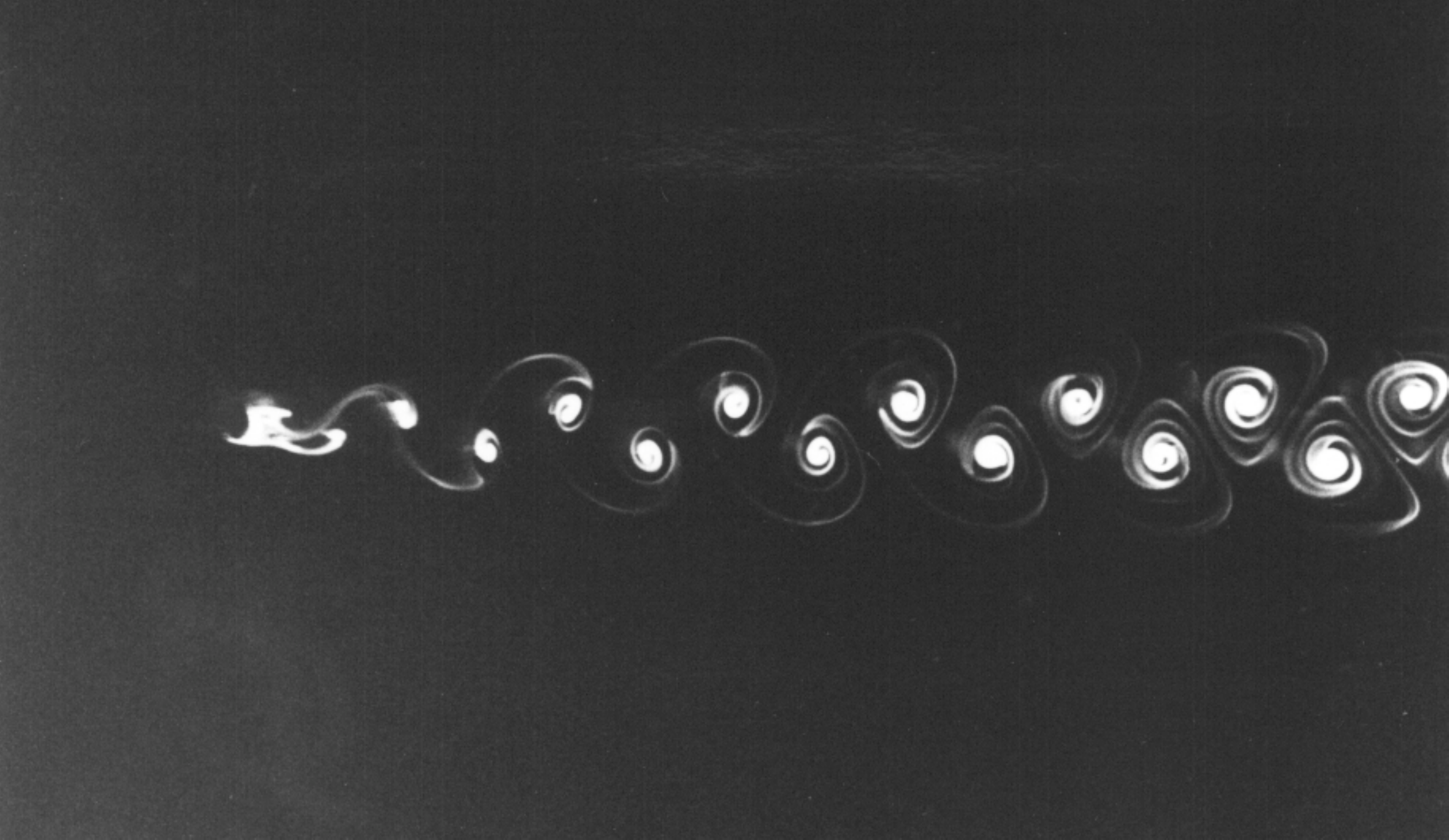}
  \caption{Flow of a classical viscous fluid past a cylinder. (a) $\Rey=41$: a steady symmetric flow behind the cylinder \cite{taneda41}. (b) $\Rey = 112$: time-dependent B\'enard--von K\'arm\'an vortex street \cite{taneda112}.}
  \label{fig:taneda-imgs}
\end{figure}
Recent experimental \cite{Tabeling,Salort}, 
numerical \cite{Nore,Kobayashi,Laurie} 
and theoretical studies \cite{Lvov}
have highlighted similarities between turbulence in quantum
fluids (e.g. superfluid helium and atomic Bose-Einstein condensates)
and turbulence in ordinary (classical) fluids \cite{Frisch}.
In particular, it is found that, in
the idealized case of homogeneous isotropic conditions away from
boundaries, the distribution of kinetic energy over the 
length scales obeys the celebrated Kolmogorov scaling of 
classical turbulence \cite{barenghi}. This similarity is remarkable,
because a superfluid has zero viscosity and vorticity is not a continuous
field but is concentrated in discrete vortex filaments of fixed 
circulation $\kappa$ proportional to Planck's constant. 
In the more realistic
presence of boundaries (such as an obstacle or confining channel
walls), superfluid hydrodynamics
is less understood, despite the large number of experiments in such scenarios. 

In a classical viscous fluid \cite{Frisch}, the prototype problem with
a boundary is the flow 
around a cylinder or a sphere (or, changing the frame of reference, 
the motion of a cylinder or a sphere in a fluid at rest).
The nature of such flow is determined by the Reynolds 
number $\Rey = vd/\nu$, where $v$ is the (assumed uniform)
flow's velocity away from the obstacle, $d$ is the obstacle's size,
and $\nu$ is the fluid's
kinematic viscosity. If $\Rey\lesssim50$, a steady symmetric 
wake forms behind the obstacle; if $10^2\lesssim\Rey\lesssim10^5$ the wake 
becomes asymmetric and time dependent, forming the famous 
B\'enard--von K\'arm\'an vortex street structure.  These cases are depicted in Figure \ref{fig:taneda-imgs}.  At even higher $\Rey$,
the flow becomes turbulent. 

What happens in a superfluid is not clear. Firstly, the superfluid has
zero viscosity ($\nu=0$) and hence $\Rey$ cannot be defined. Secondly,
experiments performed in superfluid helium confirm that the flow is affected
by the boundaries \cite{VanSciver1999,VanSciver2005}; unfortunately 
what is observed is not the flow pattern itself, but rather the
trajectories of tracer particles, whose
relation with the flow is still the subject
of investigations \cite{sergeev09}. Numerical simulations of three-dimensional (3D) superfluid flow around
an oscillating sphere performed using the vortex filament model
were not conclusive - 
quantum vortices did not appear to organise themselves
into a visible classical--like wake near the obstacle \cite{Hanninen,Fujiyama,goto08}.

The two-dimensional (2D) scenario of an obstacle moving through a superfluid offers a simplified platform to consolidate analogs and disparities between classical and quantum fluids.  In their pioneering simulations of the 2D nonlinear Schr\"odinger equation, Frisch and Pomeau \cite{frisch92} observed the formation of vortex pairs in the flow past a circular obstacle.  A more complete picture has been revealed by Sasaki {\it et al.} \cite{saito10}. Below a critical velocity (which depends on the strength \cite{jma00} and shape of the external potential), the fluid undergoes laminar flow around the obstacle.  Above this critical velocity vortices become nucleated and peel off from the moving obstacle. Two patterns are possible, depending on the size of the obstacle: vortex-antivortex pairs in either a symmetric \cite{frisch92} or asymmetric configuration (with the preference for the latter); or alternating pairs of like-signed vortices, forming a trail analogous to the Bern\'ard-von K\'arm\'an vortex street.  At higher velocities, vortex nucleation becomes highly irregular.

Recent studies of this 2D system have considered vortex emission and drag \cite{nore93,jma99,win00,huepe00}, the critical velocity \cite{zwerger00,crescimanno00,berloff2000,rica2001,pham2004}, the effect of inhomogeneous potentials \cite{win00,jackson98,fujimoto11}, the role on the obstacle parameters \cite{huepe00,aioi11}, and supersonic effects such as oblique dark solitons \cite{el06} and Cerenkov radiation \cite{carusotto06}. Neely \emph{et al.}\cite{Neely13} have shown that, within stirred annular trapping potentials, macroscopic rotation forms due to spatial clustering of vortices.

In this work we present the first clear
evidence of a classical wake in superfluid flow past an obstacle. Using
the Gross-Pitaevskii equation (GPE) for a zero-temperature Bose-Einstein condensate and an elliptical obstacle, we 
show that the interaction of  
discrete vortex singularities downstream of the
obstacle yields a flow pattern which indeed mimics classical vortex flow.

\section{Model}

We consider an atomic Bose-Einstein condensate (BEC) moving relative to a laser-induced obstacle (imposed through 
an external potential), as realized experimentally in 3D \cite{Raman,Onofrio,Inouye,Neely} and quasi-2D condensates \cite{Neely}.  This scenario closely resembles that of the classical wake-problem \cite{taneda41,taneda112}.  On a much larger scale, a similar 
3D configuration has been experimentally realized in liquid helium 
\cite{VanSciver1999,VanSciver2005}.

The BEC, assumed to be weakly-interacting and at ultracold temperature, is parameterized by its mean-field wavefunction $\Psi({\bf r},t)$, which defines the fluid number density $n({\bf r}, t)=|\Psi|^2$.  The wavefunction satisfies the non-linear Schr{\"o}dinger equation, also known as the Gross-Pitaevskii equation \cite{Pethick}, where a cubic nonlinearity arises from the mean-field potential generated by the dominant {\it s}-wave (contact) atom-atom interactions. The equation is,  
\begin{equation}
i \hbar \frac{\partial\Psi}{\partial t} = \left(-\frac{\hbar^2}{2m}\nabla^2 + V({\bf r},t) + g|\Psi|^2 - \mu \right) \Psi.
\label{eq:gpe1}
\end{equation} 
Here $g=4 \pi \hbar^2 a_{\rm s}/m$ is the interaction coefficient, with $a_{\rm s}$ being the atomic {\it s}-wave scattering length and $m$ the atomic mass, and $\mu$ is the chemical potential of the condensate.  The GPE is solved in the reference frame moving with the obstacle, at speed $v$ along $x$.  

The external potential acting on the system $V({\bf r}, t)$ is taken to be zero everywhere, i.e. a homogeneous system with uniform density $n_0$, apart from a localized repulsive potential, Gaussian in shape, which represents the obstacle.  A key feature of this work is that the obstacle is taken to be elliptical, of ellipticity $\epsilon$, with the short axis being parallel to the flow, $x$.    Such a potential, in its 2D form, can be generated via the repulsive optical dipole force from an incident blue-detuned laser beam which is moved relative to the condensate either by deflection of the beam \cite{Raman,Onofrio,Inouye} or motion of the condensate itself when offset in a harmonic trap \cite{Neely}.  While laser-induced obstacles generated to date have had a circular profile, elliptical modification of the Gaussian potential can be achieved via cylindrical focussing of the laser beam.

We express length in terms of the healing length $\xi = \hbar/\sqrt{m n_0 g}$, speed in terms of the speed of sound $c=\sqrt{n_0 g/m}$, and time in terms of $(\xi/c)$.
A detailed description of the model can be found in Appendix A.

\begin{figure}
(a) \hspace{1.8cm} (i)  $t = 1000 ~(\xi/c)$ \hspace{2.4cm} \hspace{1.8cm} (ii)   $t= 3500~ (\xi/c)$
\\
\vspace{-1.3cm}

\label{fig:dens3}\includegraphics[width=0.5\textwidth]{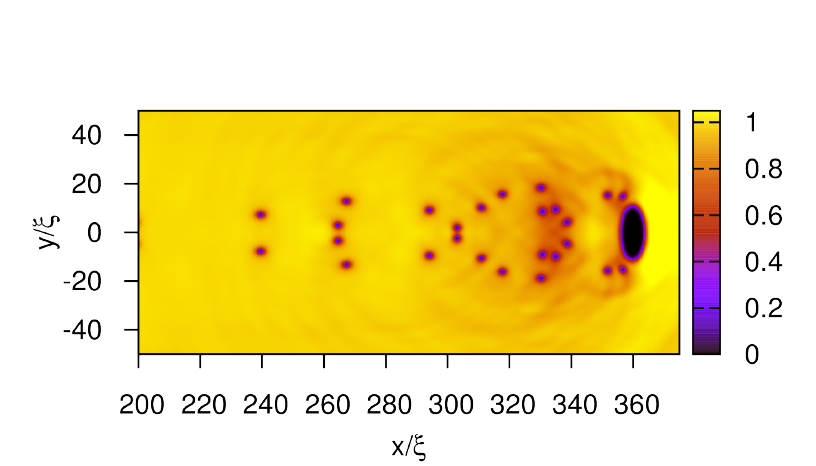}
\label{fig:dens4}\includegraphics[width=0.5\textwidth]{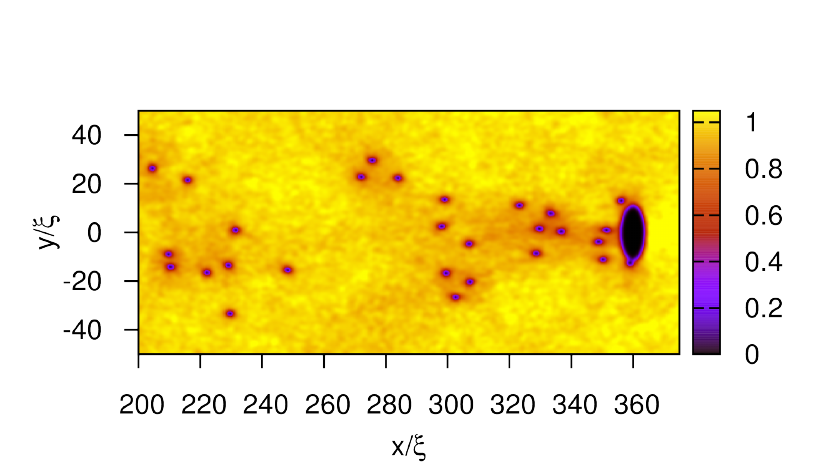}

\vspace{0.2cm}
(b)\hspace{1.5cm} (i)  $t = 0-1500 ~(\xi/c)$ \hspace{2.4cm} \hspace{0.8cm} (ii)   $t= 1600-5000~ (\xi/c)$
\\
  \includegraphics[width=0.455\textwidth]{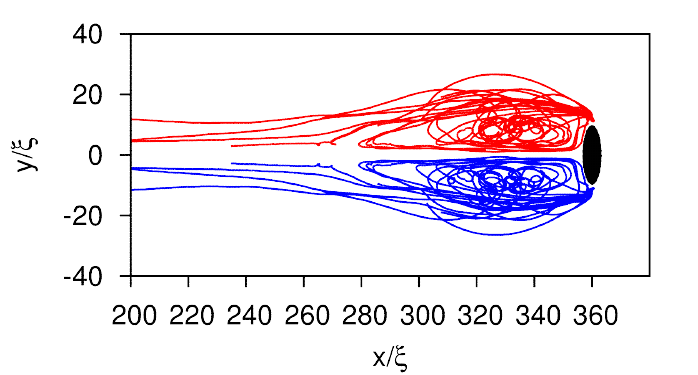}\hspace{0.045\textwidth}
  \includegraphics[width=0.455\textwidth]{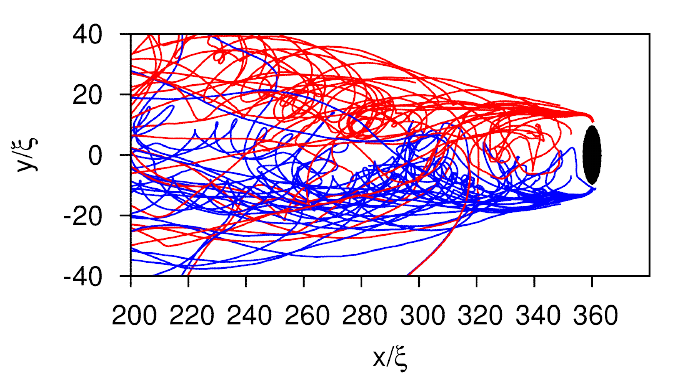}
  \caption{Snapshots showing the (a) density profile and (b) vortex trajectories during vortex shedding from an elliptical object ($\varepsilon = 3$) at (i) early times and (ii) later times.  The obstacle has speed $v=0.365c$ and size $d = 5\xi$. Red and blue lines represent vortices of oppositely quantized circulation. At early $t$, a symmetric wake similar to a classical fluid with low $Re$ forms. Symmetry breaks at $t\approx1500~(\xi/c)$ at which point vortex motion becomes disordered. In this case the initial condition is noise-free.}
  \label{fig:denstraj}
\end{figure}
\section{Results: Two-Dimensional Wakes}
We begin by exploring quantum wakes in the 2D flow of a BEC past an obstacle, according to the 2D GPE with the elliptical potential defined in Equation  (\ref{eq:potential2D}).
\subsection{Vortex emission from elliptical obstacles}

For illustrative purposes we first consider an elliptical obstacle (size $d=5\xi$, ellipticity $\epsilon=3$) moving at speed $v=0.365c$.  This speed exceeds the critical velocity for the obstacle such that quantum vortices become nucleated and trail behind to form a wake [Figure \ref{fig:denstraj}(a)].  Sound waves, also generated by the obstacle, have little effect on the vortex dynamics. At early times [Figure \ref{fig:denstraj}(a)(i)], the vortex shedding occurs through the symmetric generation of vortex-antivortex pairs, leading to a collimated and symmetric wake behind the obstacle.  This is in qualitative agreement with observations for circular obstacles  \cite{frisch92,nore93,win00,huepe00}, although, for the same obstacle velocity and size, the elliptical obstacle induces a higher frequency of vortex emission and thus a denser wake.  We examine the role of ellipticity in more detail in Sections \ref{sec:crit_vel} and \ref{sec:ellipt}.  

At later times [Figure \ref{fig:denstraj}(a)(ii)], the flow becomes asymmetric due to the known instability of symmetric wakes \cite{nore93}.  A striking pattern emerges whereby distinct clusters of co-rotating vortices (of the order of 5 vortices in each cluster) develop downstream of the obstacle.  Each cluster contains vortices of the same sign and adjacent clusters have alternating sign.  These clusters form a B\'enard--von K\'arm\'an vortex street downstream from the obstacle, confirming the intuition that a sufficiently large number of quanta of circulation reproduce classical physics.  Here, the ellipticity of the obstacle facilitates the formation of this street; the relatively high rate of vortex emission leads to a greater interaction between vortices in the wake which in turn promotes clustering.  In contrast, for a circular obstacle the symmetric wake evolves into a V-shaped wake of vortex-antivortex pairs \cite{saito10}; this because the vortex emisson rate and hence their subsequent interaction is insufficient to induce significant clustering.

The vortex trajectories provide visualisation of the time-integrated nature of the wake  [Figure \ref{fig:denstraj}(b)].   At early times (i), we see that the vortex trajectories are symmetric, forming a flow pattern in striking analog to the classical wake at low $Re$.  The generic development of vortex trajectories is as follows.  Pairs of singly-quantized vortices of opposite sign peel off from the poles of the obstacle and interact with each other as vortex-antivortex pairs.  Each pair propagates in the positive $x$ direction with approximate velocity $\hbar/(md_p)$\cite{saito10}, where $d_p$ is the pair separation \cite{Donnelly};  the pair's velocity is less than the obstacle's velocity and it drifts behind the obstacle.  As the pair moves further away from the obstacle, its separation decreases and its velocity increases, such that it begins to catch the obstacle up.  Once the pair is sufficiently close to the obstacle, it again separates and slows down, then the cycle repeats.  As more vortices are nucleated, two distinct clusters of like-circulation form.  Nucleated pairs then travel around the outside of the existing cluster before contracting, speeding up and travelling through the middle of the clusters towards the obstacle.  The clusters grow until they reach a maximum size depending on the obstacle's size and speed.  Hereafter, nucleated vortex pairs travel around the outside of the two clusters and continue travelling downstream,  becoming lost from the main wake.

\begin{figure}
\centering
\includegraphics[width=0.9\textwidth]{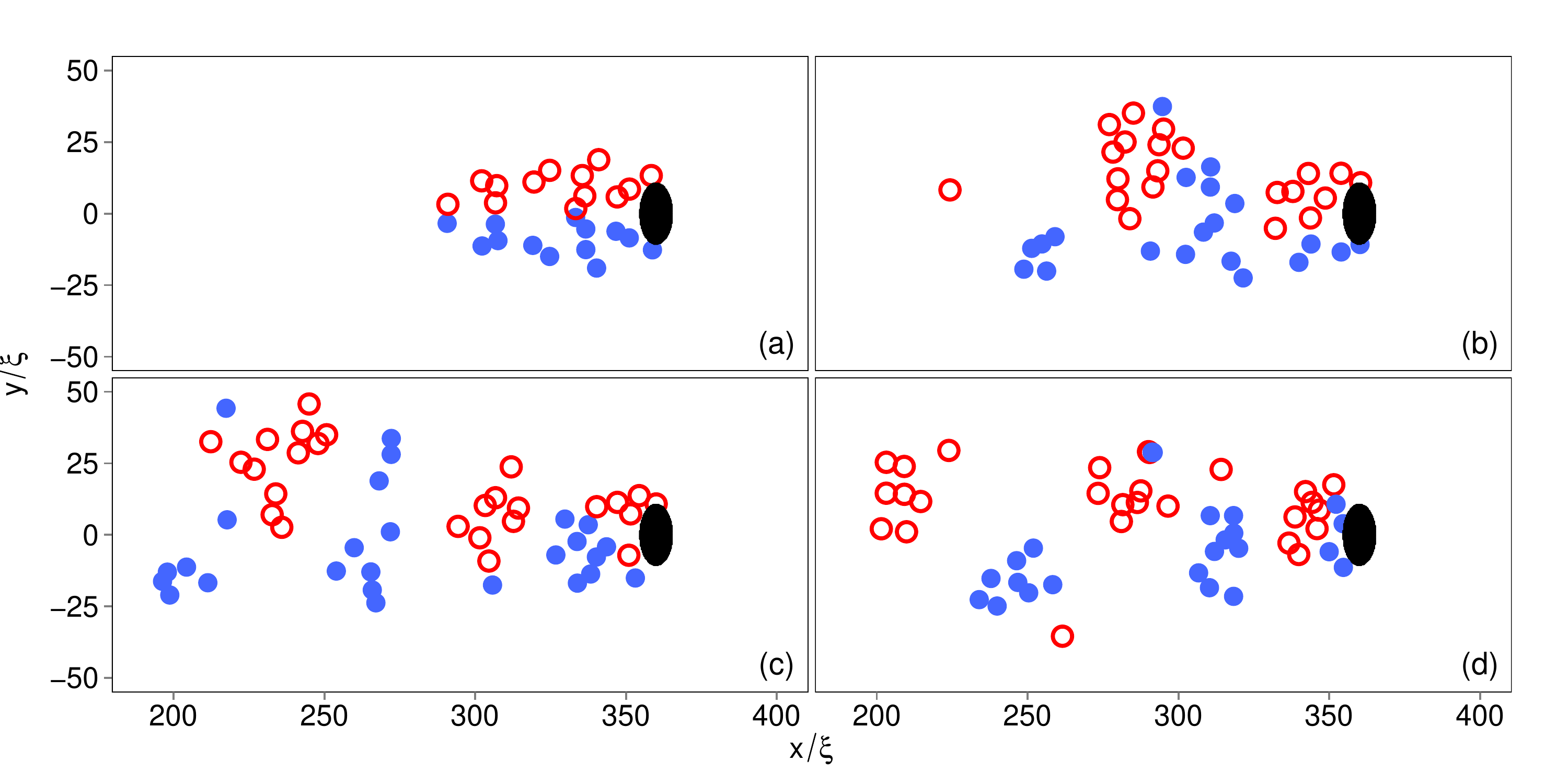}
\caption{\label{fig:progress} Snapshots of vortex locations during the motion of an elliptical object ($d=5 \xi$ and $\varepsilon=3$) at speed $v=0.52c$ in the presence of small-amplitude noise at $t=0$.  The snapshots are at times (a) $t=450$, (b) $900$, (c) $1000$ and (d) $1100~(\xi/c)$.  Red/blue circles represent vortices with quanta of circulation $+1/-1$.  
The wake forms into clusters of like-circulation that continue to be produced, in analogy to the classical B\'enard--von K\'arm\'an vortex street from a cylinder.}
\end{figure}
\subsection{Formation of the B\'enard--von K\'arm\'an vortex street}
Once the symmetry of the wake is broken, vortices no longer separate into two distinct clusters of like-circulation. Existing vortices and newly-nucleated vortices mix together behind the obstacle. However it is apparent in Figure \ref{fig:denstraj}(b)(ii) that, on average, positive vortices drift to $y>0$ while negative vortices prefer to drift to $y<0$. 

To accelerate the formation of the asymmetric wake, we subsequently seed the initial condition with noise.  Figure \ref{fig:progress} shows the vortex locations at various stages of the evolution. The initial symmetry of the wake [Figure \ref{fig:progress}(a)] breaks at $t \approx 450 (\xi/c)$, with the wake splitting into several clusters. The velocity field around the obstacle is affected: it depends on time and the distance of the nearest cluster of vortices. The obstacle no longer simultaneously produces vortex-antivortex pairs, but now generates a series of like-signed vortices.  Since like-signed vortices are known to co-rotate, these vortices group into clusters which slowly rotate.  This cluster effects the velocity field once more, causing a cluster of opposite signed vortices to be produced. This process then repeats such that clusters of like signed vortices are then produced behind the obstacle, much like vorticity in the classical vortex street behind a cylinder.  While some positive clusters contain negative vortices and vice versa, the overall pattern is still a time-dependent B\'enard--von K\'arm\'an vortex street.

For clusters consisting of pairs of vortices, it has been shown that they can survive downstream for a very long time \cite{saito10}.  However, for regimes with larger numbers of vortices in each cluster, the chaotic nature of vortex motion can cause originally tightly packed and circular clusters to easily stretch over large areas, form strange shapes, or even split into smaller clusters. Examples of this will be shown later in Figure \ref{fig:3x3grid}.

\subsection{Critical Velocity past an Elliptical Obstacle}
\label{sec:crit_vel}
\begin{figure}
\centering
\includegraphics[width=0.5\textwidth]{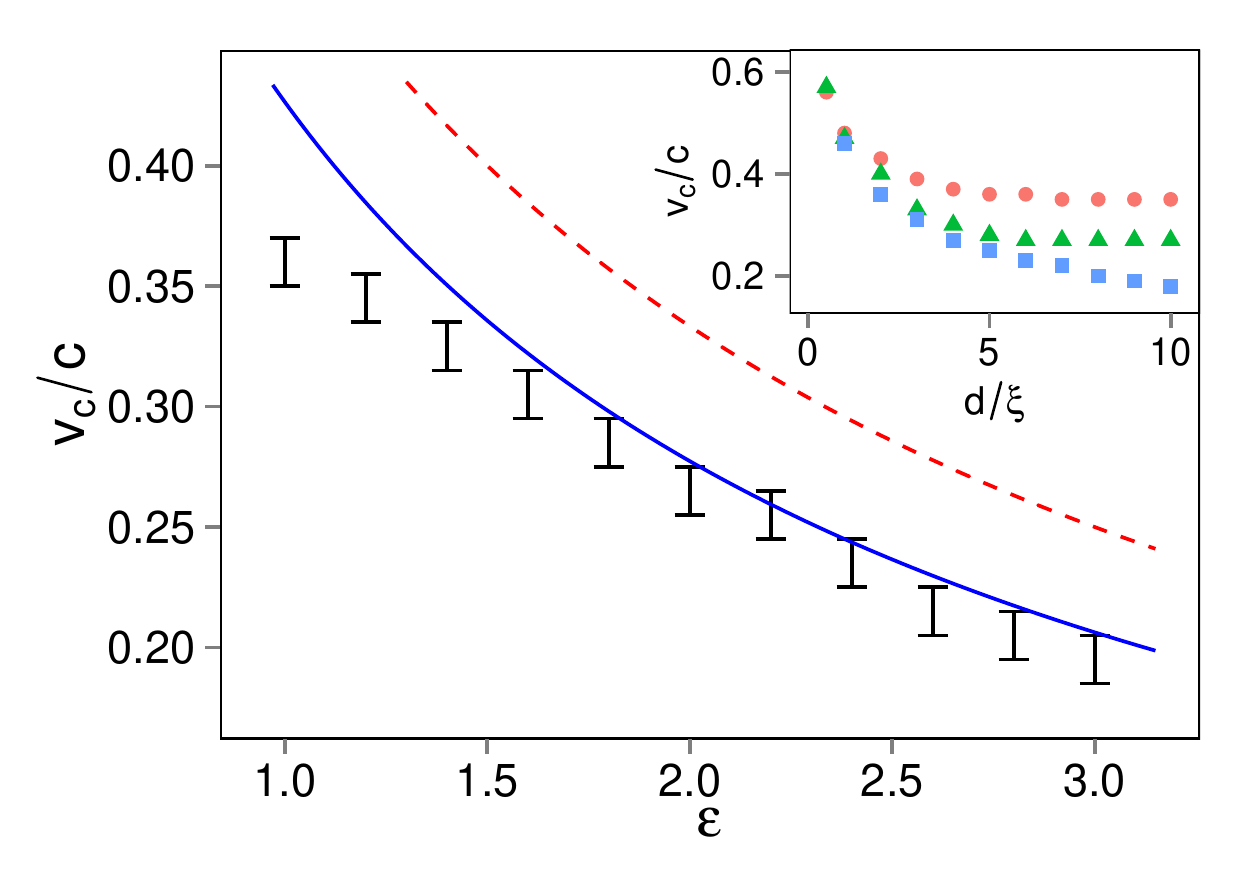} 
\caption{\label{fig:velplots}Critical velocity against obstacle ellipticity $\varepsilon$, for $d=10\xi$.  Shown are the results from the numerical simulations (black bars), Equation (\ref{eq:crit1}) (dashed red line) and Equation (\ref{eq:crit2}) (solid blue line). Inset: Critical velocity (obtained numerically) versus the obstacle width $d$, for ellipticities $\varepsilon=1$ (red circles), $\varepsilon=2$ (green triangles) and $\varepsilon=3$ (blue squares).
}
\end{figure}
{\em Elliptical} obstacles facilitate the formation of semi-classical wakes because they reduce the critical velocity and enhance the vortex shedding frequency.  Figure \ref{fig:velplots}(a) shows the critical velocity for flow past the obstacle as a function of its ellipticity, taking the obstacle to have fixed width in the $y$-direction of $d=10\xi$.  We determine the critical velocity numerically by performing simulations with flow velocities increasing in steps of $0.01$ until vortices nucleate.  
For a circular object, we find that the critical velocity is $v_{\rm c}=0.355 (\pm 0.005)c$, consistent with predictions in the Eulerian ($d \gg \xi$) limit \cite{berloff2000,rica2001,pham2004}.  As the ellipticity is increased (i.e. the obstacle becomes narrower in $x$), the critical velocity decreases.  The modification of the critical velocity is significant: if $\varepsilon=3$, $v_{\rm c}$ is more than $40\%$ smaller than that for a circular obstacle. 

The rough dependence of $v_{\rm c}$ on $\varepsilon$ can be derived as follows.   According to Landau's criterion \cite{NozieresPines},  superfluidity breaks down when the fluid velocity exceeds the critical velocity $v_{\rm Lan}=\min \left[E(p)/p\right]$, where $p$ is the momentum of elementary excitations and $E(p)$ their energy.   The weakly-interacting Bose gas has the dispersion relation $E(p)=[ngp^2/m + p^4/(4m^2)]^{1/2}$, hence $v_{\rm Lan}=c$.  If an obstacle moves through the fluid with speed $v$, the local fluid velocity at the poles exceeds $v$. Approximating the BEC as an inviscid Euler fluid undergoing potential flow about the object, then the maximum local velocity is $v_{\rm max}=(1+\varepsilon)v$ and the Landau critical velocity is (dashed red line in Figure \ref{fig:velplots}(a)),
\begin{equation}
\frac{v_{c1}}{c} = \frac{1}{1+\varepsilon}.
\label{eq:crit1}
\end{equation}

While this result assumes constant density, a first order correction can be made by using Bernoulli's theorem to model the reduction in local density near the obstacle (due to the enhanced local fluid velocity) which in turn reduces the local speed of sound $c(x,y)=\sqrt{n(x,y)g/m}$ \cite{win01}.  This then leads to the modified result,
\begin{equation}
\frac{v_{c2}}{c} = \left [\frac{3}{2}(1+\varepsilon)^2 - \frac{1}{2}\right]^{-\frac{1}{2}}.
\label{eq:crit2}
\end{equation}
This relation (solid blue line in Figure \ref{fig:velplots}) gives good agreement with the computed values of $v_{\rm c}$.  The deviation for $\varepsilon \sim 1$ has been noted elsewhere \cite{rica2001}, and can be remedied using higher order corrections.

From studies on circular objects, it is known that $v_{\rm c}$ depends on the obstacle's shape at small diameters, where boundary layer effects are significant; $v_{\rm c}$ approaches the ``Eulerian" value only for large diameters $d \gg \xi$ \cite{huepe00,rica2001}.  The variation of $v_{\rm c}$ with the obstacle width $d$ is shown in Figure \ref{fig:velplots} (inset).  For $d=10\xi$, the critical velocity effectively reaches its asymptotic value, while at smaller widths, it is much larger.    

\subsection{Role of Obstacle Size and Ellipticity on the Wake}
\label{sec:ellipt}
\begin{figure}
\includegraphics[width=\textwidth]{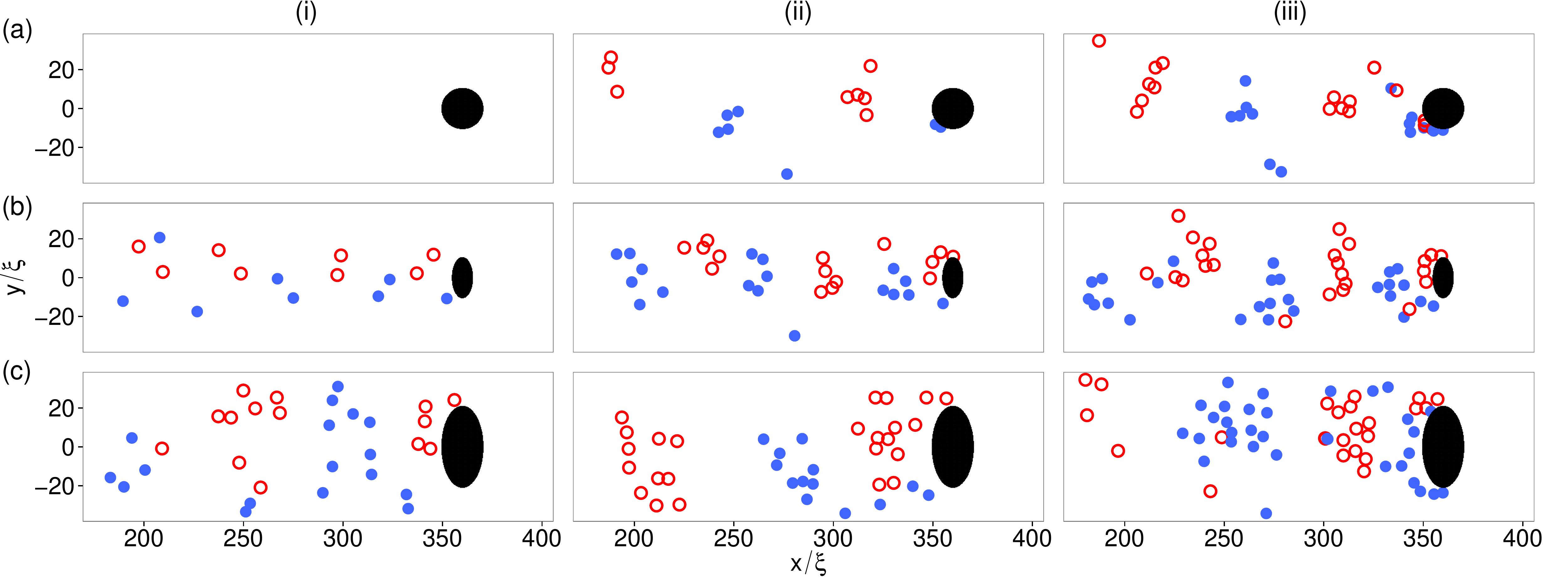}
\caption{\label{fig:3x3grid}Snapshots of the vortex positions for various obstacle parameters, at $t=2000~(\xi/c)$. Shown are obstacles corresponding to (a) $\varepsilon=1$ and $d=5\xi$, (b) $\varepsilon=2$ and $d=5\xi$, and (c) $\varepsilon=2$ and $d=10\xi$, at the velocities (i) $v=0.32c$, (ii) $v=0.40c$, and (iii) $v=0.48c$.  Red/blue circles represent vortices with quanta of circulation $+1/-1$.}
\end{figure}
During the initial symmetric phase of vortex nucleation, the wakes generated by the obstacle have the same qualitative structure shown in Figure \ref{fig:denstraj}(b) (i).  However, once the wake becomes asymmetric, the nature of the clusters that form are highly dependent on the velocity and shape of the obstacle. Figure \ref{fig:3x3grid} shows wakes generated for various obstacle parameters, all captured at the same time $t=2000~(\xi/c)$. We find that any increase of size, ellipticity or velocity of the obstacle increases the number of vortices in the wake's clusters.

The shedding frequency of vortices increases with the velocity of the flow \cite{jma99}. For an elliptical obstacle, the combination of a reduced critical velocity and increased local velocity around the obstacle has the effect of increasing the shedding frequency with $\varepsilon$ and $d$. The overall result is that, when increasing any of $v$, $\varepsilon$ or $d$, more vortices are nucleated in a given time period, causing the cluster size to increase. This increase in cluster size is investigated in the next section.

\subsection{Vortex Clustering}
We have shown that the B\'enard--von K\'arm\'an vortex street forms through the clustering of like-signed vortices. Methods of quantifying the clustering of vortices in quantum fluids have been explored in the literature \cite{white12,and13,bagg12}. Here we utilize the algorithm of Reeves \emph{et. al.} \cite{and13} to identify clusters.

\begin{figure}
\centering
\includegraphics[width=0.5\textwidth]{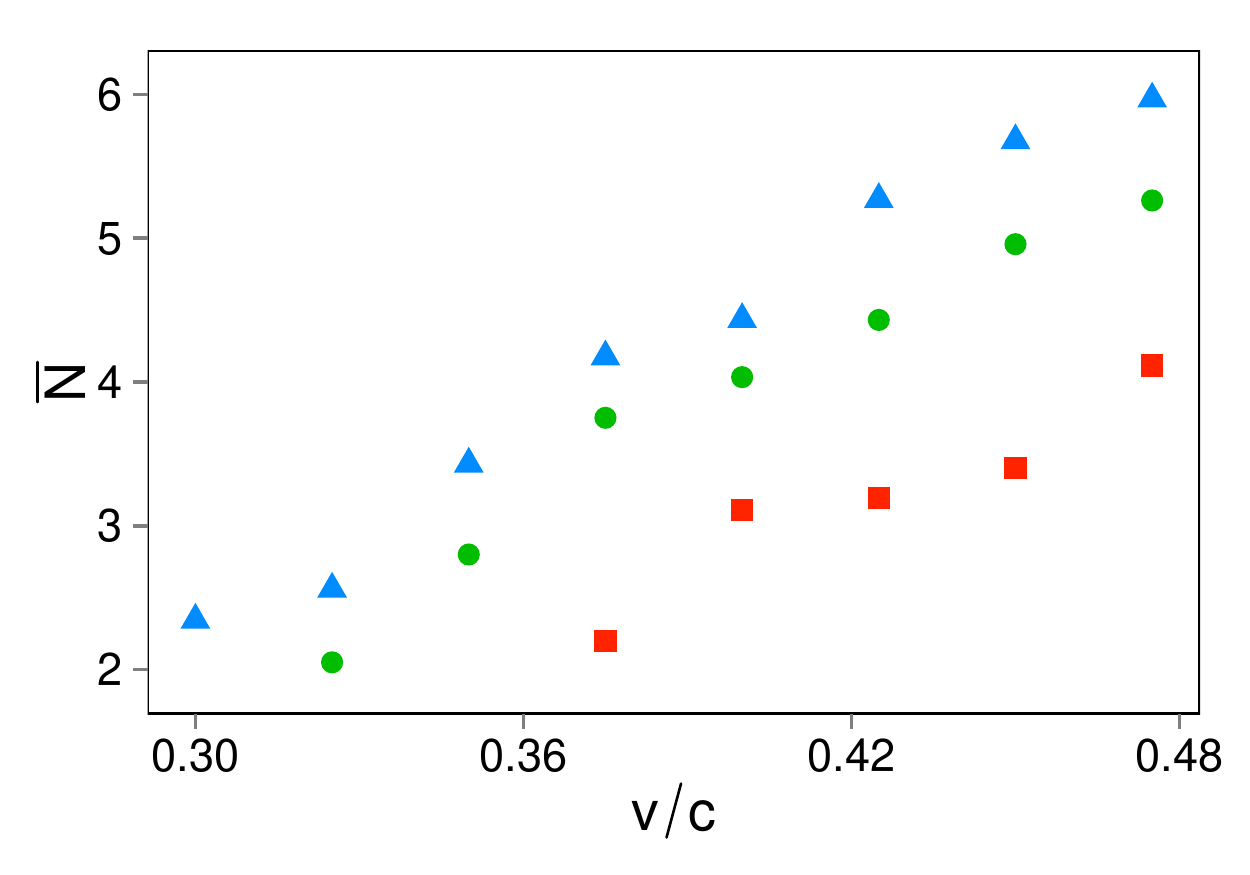}
\caption{\label{fig:Svslocal2} Average number of vortices in the clusters as a function of the obstacle velocity $v$.  Shown are cases with $\varepsilon=1$ (red squares),  $\varepsilon=2$ (green circles) and $\varepsilon=3$ (blue triangles).  All cases feature $d=5\xi$.}
\end{figure}

Firstly we record the number of clusters $N_c$ and the number of vortices in each cluster $N_i$, where $i$ is the cluster index. Then we determine the average number of vortices in the clusters, $\bar{N} = (1/N_c) \sum_{i=1}^{i=N_c} N_i$ as a function of obstacle velocity $v$ for three ellipticities $\varepsilon=1, 2$ and $3$, at times $t=500(\xi/c),510(\xi/c),\ldots,2500(\xi/c)$. The results, plotted in Figure \ref{fig:Svslocal2}, show that increasing $v$ (above the critical velocity) causes $\bar{N}$ to increase and that, at fixed $v$, $\bar{N}$ increases with $\varepsilon$. We attribute this to an object with a larger $\varepsilon$ having a lower critical velocity and producing more vortices at the same $v$.  This result explains why an elliptical obstacle efficiently generates a semi-classical wake composed of large vortex clusters. We also find that for all values of $\varepsilon$, a large obstacle velocity ($v\gtrsim0.6$) causes vortices to nucleate non-periodically, inducing an irregular flow without a visible B\'enard--von K\'arm\'an vortex street configuration, in agreement with previous simulations with circular obstacles of smaller diameter \cite{saito10}.

\section{Results: Three-Dimensional Wakes}
We now generalize our results to 3D by considering quantum wakes in three-dimensional flow past a localized obstacle, as simulated via the 3D GPE with the 3D obstacle potential of Equation (\ref{eq:potential3D}).  Our results will confirm that the features observed in 2D wakes also arise in the 3D setting.  A comprehensive study of the parameter space is, however, not tractable in 3D due to the computational intensity of the 3D simulations.

\begin{figure}
\centering
\includegraphics[width=\textwidth]{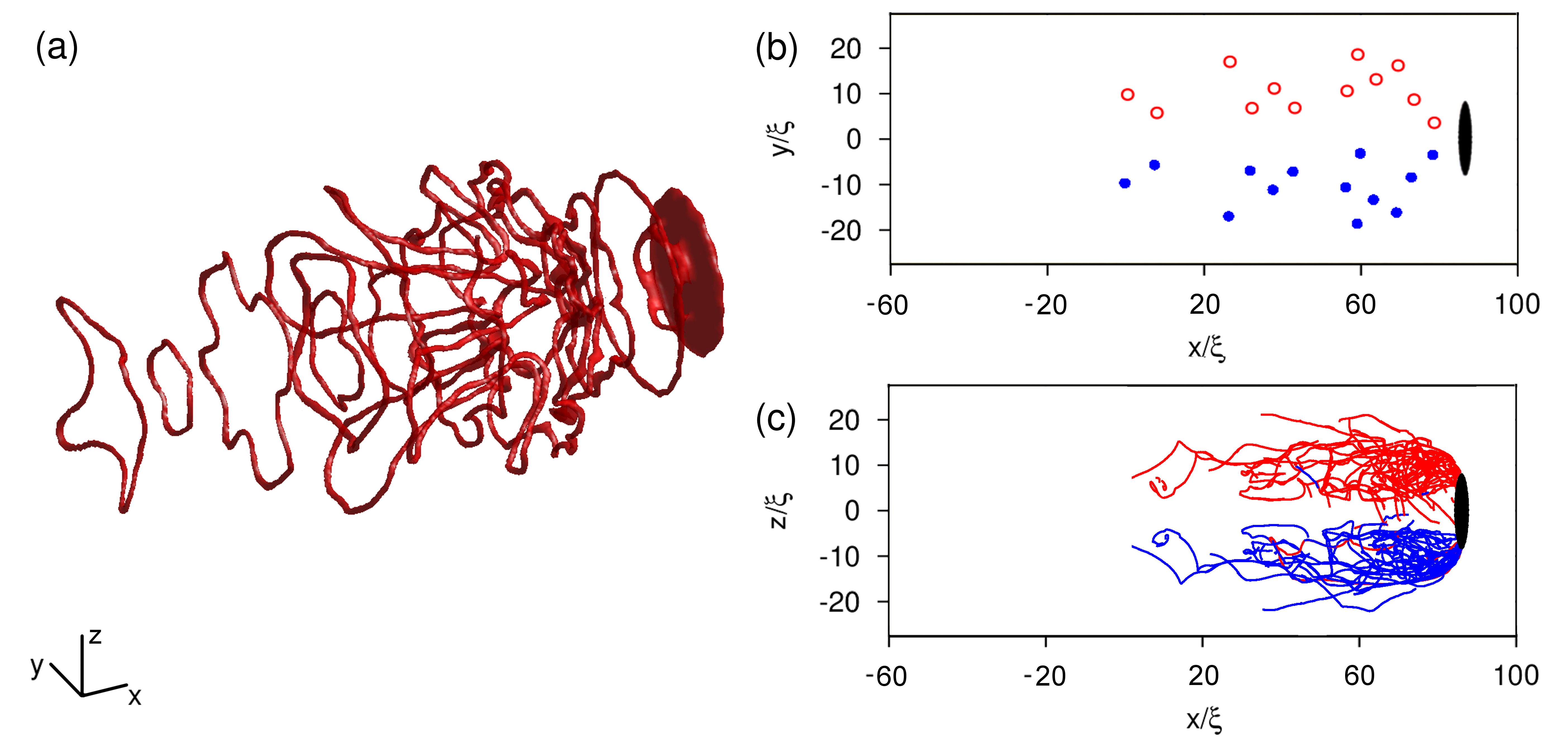}
\caption{\label{fig:3d1} Symmetric wake in 3D at $t=450~(\xi/c)$ for an elliptical obstacle ($d=5\xi$ and $\epsilon=5$) moving at $v=0.6c$.  (a) Isosurface plot of low density, over a range $[0,100]$ in $x$ and $[-25,25]$ in $y$ and $z$. (b) Vortex locations in the $xy$ plane.  (c) Vortex trajectories in the $xz$ plane.  Here (b) and (c) show opposing circulation in red and blue.}
\end{figure}
\subsection{Symmetric Wakes} 
For a spherical ($\varepsilon=1$) object with $d=5\xi$, we find that the critical velocity is $v_c=0.455\pm 0.05 c$, consistent with $v_c=0.55c$ reported in the Eulerian limit ($d \gg \xi$) \cite{win01,winiecki99}.  Making the obstacle ellipsoidal, with the short direction parallel to the flow, reduces the critical velocity, in parallel with our 2D observations.  For example, for $\varepsilon=5$, the critical velocity is reduced to $v_c=0.315 \pm 0.05c$.  Figure \ref{fig:3d1}(a) shows the 3D wake generated past this ellipsoidal obstacle ($d=5\xi$ and $\varepsilon = 5$) when moving at super-critical speed $v=0.6c$.  Vortex rings, the 3D analog of vortex-antivortex pairs, are ejected at high frequency (due to the obstacles high ellipticity) in the direction of the flow.  At early times ($t=450~(\xi/c)$ in this case) the vortex configuration maintains cylindrical symmetry about the obstacle's axis, as is clearly visible in the $xy$ and $xz$ planes in Figure \ref{fig:3d1}(b) and (c).  As the vortex rings move downstream they shrink and speed up, returning to the object, sometimes passing through other vortex rings. A similar behaviour is observed \cite{wacks} in the evolution of toroidal bundles of many coaxial vortex rings which leapfrog around each other.  Occasionally a ring will escape this cycle and fall downstream.  These behaviors conspire to form an organized symmetric wake behind the obstacle,  the 3D analog of our 2D observations.

\subsection{Asymmetric Wakes}
\begin{figure}
\centering
\includegraphics[width=\textwidth]{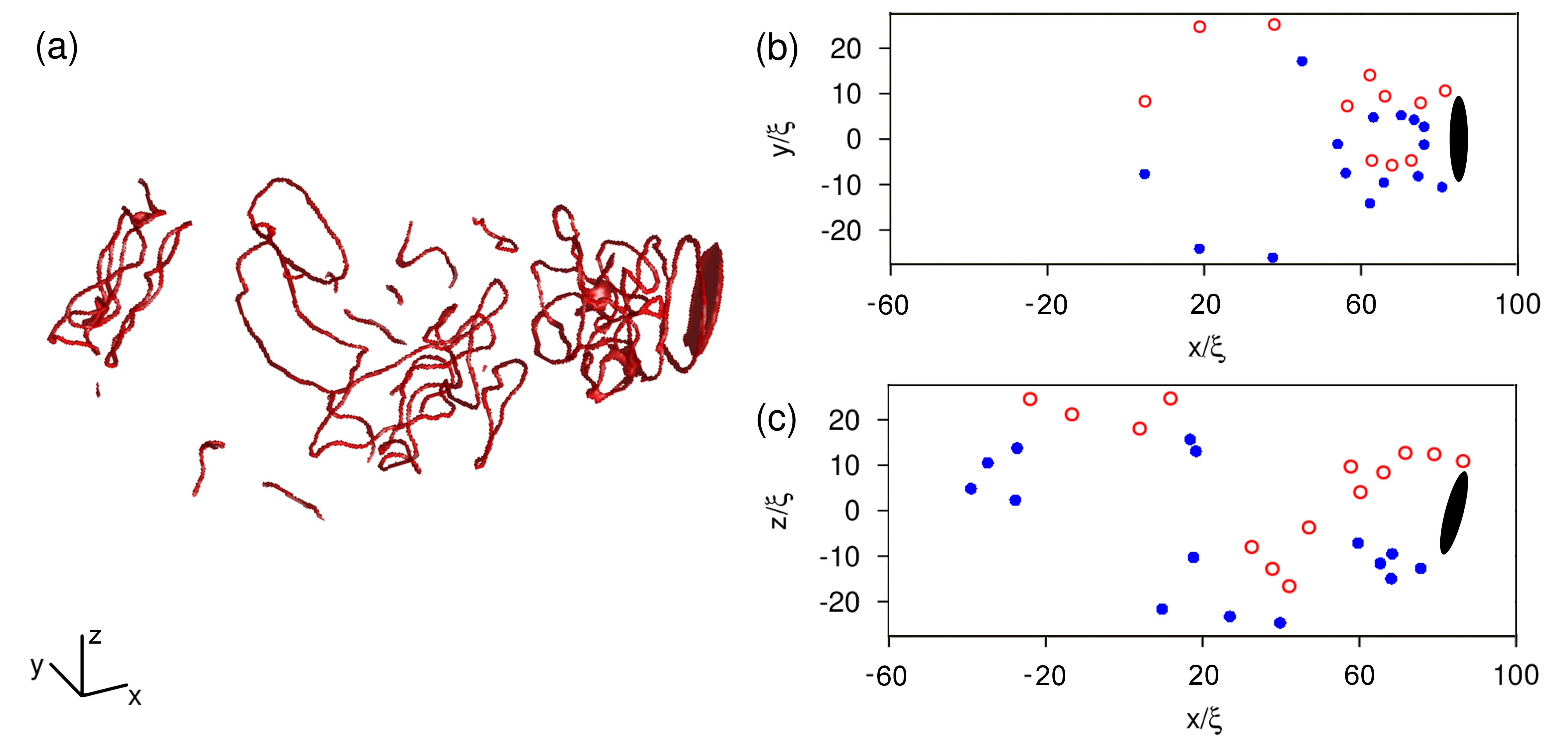}
\caption{\label{fig:3d4} Asymmetric wake in 3D at $t=340~(\xi/c)$ for an elliptical obstacle ($d=5\xi$ and $\epsilon=5$) moving at $v=0.6c$. (a) Isosurface plot of low density, over a range $[-60,100]$ in $x$ and $[-25,25]$ in $y$ and $z$. (b) Vortex locations in the $xy$ plane.  (c) Vortex locations in the $xz$ plane.  Here (b) and (c) show opposing circulation in red and blue.}
\end{figure}

We break the cylindrical symmetry of the system by tilting the obstacle by a small angle in the $xz$ plane.  The vortex rings, illustrated in Figure \ref{fig:3d4}, now become ejected and evolve asymmetrically; Kelvin waves and reconnections occur, forming an apparently disordered tangle of vortices behind the obstacle.  Due to the manner in which symmetry is broken, the wake remains approximately symmetric in the $xy$ plane, as evident in Figure \ref{fig:3d4} (b).  However, unlike in Figure \ref{fig:3d4}, the vortices do not self organise into two clusters of alternate circulation. This is due to the vortex rings interacting, reconnecting and shifting out of the plane (which manifests in 2D as two alternate-sign vortices approaching one another).

However, in the $xz$ plane (Figure \ref{fig:3d4} (c)), symmetry is broken. Due to the relatively high frequency of vortex nucleation and relatively low flow speed, like signed vortices cluster together as they are ejected by the obstacle, much like the 2D solutions seen in earlier sections.  Downstream the tangle may shift both across or out of the plane. In 2D, although this manifests as a shift in location of the vortex clusters, the clusters largely remain rather than forming dipoles. 

\section{Conclusion}
We have shown that the motion of an obstacle in a Bose-Einstein condensate produces classical-like wakes consisting of quantum vortices of the same polarity.  This is consistently observed in both two- and three-dimensional scenarios.  The key ingredient to produce classical-like wakes - that vortices are generated at a sufficiently high rate that they undergo strong interactions with their neighbours  (rather than being swept away) - is that the obstacle is elliptical, which reduces the critical velocity for vortex nucleation.  Symmetric wakes resemble those observed in classical flow at low $\Rey$.  These are unstable, forming time-dependent asymmetric structures similar to the B\'enard--von K\'arm\'an vortex street of classical fluid dynamics. Vortex singularities
in the inviscid superfluid thus mimic classical vortex patterns typical of viscous flows.  The effects which we describe (dependence of the critical velocity and cluster size on the obstacle's size, velocity and ellipticity) can be experimentally studied in atomic Bose-Einstein condensates using moving laser-induced potentials. They are also relevant to the motion of objects (such as vibrating wires, grids and forks) in superfluid helium, as the obstacle's ellipticity plays a role which is analogous to rough boundaries \cite{blaz08,brad05}.  

\ack
This work made use of the facilities of N8 HPC provided and funded by the N8 consortium and EPSRC (Grant No.EP/K000225/1). These facilities are co-ordinated by the Universities of Leeds and Manchester. We also thank the EPSRC (Grant No.EP/I019413/1) for financial support.
\\
\appendix
\section{Model of 2D and 3D BEC with Gaussian Potentials}

In our 3D simulations, we solve the 3D GPE of Equation (1), where the localized 3D obstacle is modelled via a repulsive ellipsoidal Gaussian potential,
\begin{equation}
V({\bf r}, t)=V_0 \exp \left( -\frac{\varepsilon^2(x-x_0-vt)^2}{d^2} -\frac{(y-y_0)^2}{d^2}-\frac{(z-z_0)^2}{d^2}\right),
\label{eq:potential3D}
\end{equation}
where  $V_0$ is its (constant) amplitude, $d$ its width in the $y$ and $z$ directions, and $(x_0,y_0,z_0)$ its initial coordinates.  The GPE is transformed into the reference frame moving with the obstacle (in $x$) via the addition of the Galilean term $ i\hbar v\frac{\partial}{\partial x} \Psi$ to the right-hand side of the GPE (\ref{eq:gpe1}), where $v$ is the frame velocity.  

To effectively reduce the system to two-dimensions, the BEC is assumed to be confined by a harmonic trapping potential in the axial ($z$) direction, $V(z)=\frac{1}{2}m \omega_z^2 z^2$, where $m$ is the atomic mass.  For sufficiently strong trapping, which requires $\hbar \omega_z \gg \mu$, where $\mu$ is the chemical potential of the 3D condensate, the axial wavefunction becomes ``frozen" into the time-independent harmonic oscillator ground state $\pi^{-1/4} l_z^{-1/2} \exp\left(-z^2/2l_z^2\right)$, where $l_z=\sqrt{\hbar/m \omega_z}$ is the axial harmonic oscillator length. Under these conditions, the condensate becomes effectively two-dimensional, as achieved experimentally \cite{Gorlitz}.  It is then described by an 2D GPE, corresponding to Equation (\ref{eq:gpe1}) with $g \rightarrow g/\sqrt{2\pi}l_z$ and where $\Psi$, ${\bf r}$, $V$ and $n$ become two-dimensional quantities.  

In 2D, we model the obstacle via a moving repulsive Gaussian potential of the form,
\begin{equation}
V({\bf r}, t)=V_0 \exp \left( -\frac{\varepsilon^2(x-x_0-vt)^2}{d^2} -\frac{(y-y_0)^2}{d^2}\right).
\label{eq:potential2D}
\end{equation}

The 3D (2D) system is simulated using the 4th-order Runge-Kutta method under periodic boundary conditions on a $400 \times 150 \times 150$ ($2048 \times 512$)  grid with uniform spacing $\Delta=0.4\xi$.  The obstacle is positioned upstream in the box to enable a long simulation time before vortices recycle through the periodic box.  We have verified that our simulations are well-converged, that is, increasing the grid resolution has negligible effect on the results.
The computational box is sufficiently large that the boundary conditions do not play a role in vortex shedding.  The initial condition is the stationary state of the GPE (including obstacle potential) with $v=0$ (as determined by the imaginary time convergence method). Setting $V_0=100~\mu$ throughout, the external potential closely approximates an impenetrable obstacle. Unless stated otherwise, a small amount of noise is added to the initial condition to break symmetry: a random number between $-0.0005$ and $0.0005$ is added to both the real and imaginary parts of the initial wavefunction. 

To minimize initial generation of waves, $v$ is ramped up in time along a hyperbolic tangent curve, from $v=0$ at $t=0$ to its terminal value at around $t\approx100~(\xi/c)$. During the evolution, the vortices are located (and their circulation evaluated) using an algorithm based on those of references \cite{fos10} and \cite{white12}.
\\
\section*{References}

\begin{thebibliography}{99}
	\bibitem{taneda41}
	Taneda S 1956 {\it J. Phys. Soc. Jpn.} {\bf 11} No. 3, 302
	\bibitem{taneda112}
	Taneda S 1981 {\it J. Phys. Soc. Jpn.} {\bf 50} No. 4, 1398
	\bibitem{Tabeling}
	Maurer J and Tabeling P 1998 {\it Europhys. Lett.} {\bf 43} 29
	\bibitem{Salort}
	Salort J \emph{et al} 2010 {\it Phys. Fluids} {\bf 22} 125102
	\bibitem{Nore}
	Nore C, Abid M and Brachet M E 1997 {\it Phys. Rev. Lett.} {\bf 78} 3896.
	\bibitem{Kobayashi}
	Kobayashi M and Tsubota M 2005 {\it Phys. Rev. Lett.} {\bf 94} 065302
	\bibitem{Laurie}
	Baggaley A W, Laurie J and Barenghi C F 2012 {\it Phys. Rev. Lett.} {\bf 109} 205304
	\bibitem{Lvov}
	L'vov V S, Nazarenko S V and Volovik G E 2004 {\it JETP Letters} {\bf 80} 479
	\bibitem{Frisch}
	Frisch U 1995 {\it Turbulence. The legacy of A.N. Kolmogorov} (Cambridge: Cambridge University Press)
	\bibitem{barenghi}
	Barenghi C F, L'vov V and Roche P E (in press) 2014 {\it Proc. Nat. Acad. Science}
	\bibitem{VanSciver1999}
	Smith M R, Hilton D K and Van Sciver S 1999 {\it Phys. Fluids} {\bf 11} 1
	\bibitem{VanSciver2005}
	Zhang T and Van Sciver S 2005 {\it Nature Physics} {\bf 1} 36
	\bibitem{sergeev09}
	Sergeev Y A and Barenghi C F 2009 {\it J. Low Temp. Phys.} {\bf 157} 429
	\bibitem{Hanninen}
	H\"anninen R, Tsubota M and Vinen W F 2007 {\it Phys. Rev.} B {\bf 75} 064502
	\bibitem{Fujiyama}
	Fujiyama S and Tsubota M 2009 {\it Phys. Rev.} B {\bf 79} 094513
	\bibitem{goto08}
	Goto R \emph{et al.} 2008 {\it Phys. Rev. Lett.} {\bf 100} 045301
	\bibitem{frisch92}
	Frisch T, Pomeau Y and Rica S 1992 {\it Phys. Rev. Lett.} {\bf 69} 1644
	\bibitem{saito10}
	Sasaki K, Suzuki N and Saito H 2010 {\it Phys. Rev. Lett.} {\bf 104} 150404
	\bibitem{jma00}
	Jackson B, McCann J F and Adams C S 2000 {\it Phys. Rev.} A {\bf 61} 051603
	\bibitem{nore93}
	Nore C, Brachet M E and Fauve S 1993 {\it Physica} D {\bf 65} 154
	\bibitem{jma99}
	Winiecki T, McCann J F and Adams C S 1999 {\it Phys. Rev. Lett.} {\bf 82} 5186
	\bibitem{win00}
	Winiecki T, Jackson B, McCann J F and Adams C S 2000 {\it J. Phys.} B {\bf 33} 4069
	\bibitem{huepe00}
	Huepe C and Brachet M E 2000 {\it Physica D} {\bf 140} 126
	\bibitem{zwerger00}
	Stie{\ss}berger J S and Zwerger W 2000 {\it Phys. Rev.} A {\bf 62} 061601(R)
	\bibitem{crescimanno00}
	Crescimanno M, Koay C G, Peterson R and Walsworth R 2000 {\it Phys. Rev.} A {\bf 62} 063612
	\bibitem{berloff2000}
	Berloff N G and Roberts P H 2000 {\it J. Phys.} A {\bf 33} 4025
	\bibitem{rica2001}
	Rica S 2001 {\it Physica} D {\bf 148} 221
	\bibitem{pham2004}
	Pham C T, Nore C and Brachet M E 2004 {\it Comptes Rendus Physique} {\bf 5} 3 
	\bibitem{jackson98}
	Jackson B, McCann J F and Adams C S 1998 {\it Phys. Rev. Lett.} {\bf 80} 3903
	\bibitem{fujimoto11}
	Fujimoto K and Tsubota M 2011 {\it Phys. Rev.} A {\bf 83} 053609
	\bibitem{aioi11}
	Aioi T, Kaadukura T, Kishimoto T and Saito H 2011 {\it Phys. Rev.} X {\bf 1} 021003
	\bibitem{el06}
	El G A, Gammal A and Kamchatnov A M 2006 {\it Phys. Rev. Lett.} {\bf 97} 180405
	\bibitem{carusotto06}
	Carusotto I, Hu S X, Collins L A and Smerzi A 2006 {\it Phys. Rev. Lett.} {\bf 97} 260403
	\bibitem{Neely13}
	Neely T W {\it et al} 2013 {\it Phys. Rev. Lett.} {\bf 111} 235301
	\bibitem{Raman}
	Raman C {\it et al} 1999 {\it Phys. Rev. Lett.} {\bf 83} 2502
	\bibitem{Onofrio}
	Onofrio R {\it et al} 2000 {\it Phys. Rev. Lett.} {\bf 85}, 2228
	\bibitem{Inouye}
	Inouye S {\it et al} 2001 {\it Phys. Rev. Lett.} {\bf 87} 080402
	\bibitem{Neely}
	Neely T W {\it et al} 2010 {\it Phys. Rev. Lett.} {\bf 104} 160401
	\bibitem{Pethick}
	Pethick C J and Smith H 2002 {\it Bose-Einstein Condensation in Dilute Gases} (Cambridge: Cambrige University Press)
	\bibitem{Gorlitz}
	G{\"o}rlitz A {\it et al} 2001 {\it Phys. Rev. Lett.} {\bf 87} 130402
	\bibitem{fos10}
	Foster C J, Blakie P B and Davis M J 2010 {\it Phys. Rev.} A {\bf 81} 023623
	\bibitem{white12}
	White A C, Barenghi C F and Proukakis N P 2012 {\it Phys. Rev.} A {\bf 86} 013635
	\bibitem{Donnelly}
	Donnelly R J 1991 {\it Quantized Vortices in Helium II} (Cambridge: Cambridge University Press)
	\bibitem{NozieresPines}
	Nozi\`eres P and Pines D 1990 {\it The Theory of Quantum Liquids II} (Redwood City: Addison-Wesley)
	\bibitem{win01}
	Winiecki T 2001 {\it Numerical Studies of Superfluids and Superconductors} Ph.D. thesis (The University of Durham)
	\bibitem{and13}
	Reeves M T, Billam T P, Anderson B P and Bradley A S 2013 {\it Phys. Rev. Lett.} {\bf 110} 104501
	\bibitem{bagg12}
	Baggaley A W, Barenghi C F, Shukurov A and Sergeev Y A 2012 {\it Europhys. Lett.} {\bf 98} 26002
	\bibitem{blaz08}
	Blazkova M, Schmoranzer D, Skrbek L and Vinen W F 2009 {\it Phys. Rev.} B {\bf 79} 054522
	\bibitem{brad05}
	Bradley D I {\it et al} 2005 {\it Phys. Rev Lett.} {\bf 95} 035302
	\bibitem{wacks}
	Wacks D H, Baggaley A W and Barenghi C F 2014 {\it Physics of Fluids} {\bf 26} 027102
	\bibitem{winiecki99}
	Winiecki T, McCann J F and Adams C S 1999 {\it Europhy. Lett.} {\bf 48} 475 
\end{thebibliography}

\end{document}